\documentclass[conference]{IEEEtran}
\IEEEoverridecommandlockouts
\usepackage{glossaries}
\usepackage[normalem]{ulem}
\usepackage{booktabs}
\usepackage{tabularx}
\useunder{\uline}{\ul}{}
\usepackage{cite}
\usepackage{amsmath,amssymb,amsfonts}
\usepackage{algorithmic}
\usepackage{algorithm}  
\usepackage{graphicx}
\usepackage{url}
\usepackage{textcomp}
\usepackage{xcolor}
\usepackage[perpage]{footmisc}
\usepackage{float} 
\usepackage{bm}
\usepackage{makecell}
\usepackage{multirow}
\usepackage{array}
\usepackage{booktabs}
\usepackage{multirow}
\usepackage{fancyhdr} 
\usepackage{color}
\usepackage{tabu}
\usepackage{ulem}
\usepackage{nomencl}
\makenomenclature
\usepackage{etoolbox}
\usepackage{graphicx}
\usepackage{subfigure}
\usepackage[cal=boondoxo]{mathalfa}
\newcolumntype{I}{!{\vrule width 1.3pt}}
\newlength\savedwidth

\newlength\savewidth

\def\BibTeX{{\rm B\kern-.05em{\sc i\kern-.025em b}\kern-.08em
    T\kern-.1667em\lower.7ex\hbox{E}\kern-.125emX}}

\newacronym{6G}{6G}{Sixth-Generation}
\newacronym{IDN}{IDN}{Intent-Driven Network}
\newacronym{NLP}{NLP}{Natural Language Processing}
\newacronym{RAN}{RAN}{Radio Access Network}
\newacronym{GenAI}{GenAI}{Generative Artificial Intelligence}
\newacronym{LLM}{LLM}{Large Language Model}
\newacronym{SLA}{SLA}{Service Level Agreement}
\newacronym{API}{API}{Application Programming Interface}
\newacronym{QLoRA}{QLoRA}{Quantized Low Rank Adaptation} 
\newacronym{5G}{5G}{Fifth-Generation}
\newacronym{DRL}{DRL}{Deep Reinforcement Learning}
\newacronym{MARL}{MARL}{Multi-Agent RL}
\newacronym{HRL}{HRL}{Hierarchical RL}
\newacronym{QoS}{QoS}{Quality-of-Service}
\newacronym{RAG}{RAG}{Retrieval Augmented Generation}
\newacronym{RLHF}{RLHF}{Reinforcement Learning with Human Feedback}
\newacronym{KPI}{KPI}{Key Performance Indicator}
\newacronym{RL}{RL}{Reinforcement Learning}
\newacronym{GPTQ}{GPTQ}{Generative Pre-trained Transformer-Quantization}
\newacronym{HDMGA}{HDMGA}{Hierarchical Decision Mamba with Goal Awareness}
\newacronym{DT}{DT}{Decision Transformer}
\newacronym{SSSM}{SSSM}{Selective-State Space Modeling}
\newacronym{SSM}{SSM}{State Space Model}
\newacronym{HDTGA}{HDTGA}{Hierarchical Decision Transformer with Goal Awareness}
\newacronym{RAT}{RAT}{Radio Access Technology}
\newacronym{NR}{NR}{New Radio}
\newacronym{LTE}{LTE}{Long-Term Evolution}
\newacronym{URLLC}{URLLC}{Ultra-Reliable Low-Latency Communication}
\newacronym{IRC}{IRC}{Intelligent RAN Controller}
\newacronym{DQN}{DQN}{Deep-Q-Network}
\newacronym{TTI}{TTI}{Transmission Time Interval}
\newacronym{PEFT}{PEFT}{Parameter Efficient Fine-Tuning}
\newacronym{CDL}{CDL}{Clustered Delay Line}
\newacronym{O-CU}{O-CU}{O-RAN Centralized Unit}
\newacronym{O-DU}{O-DU}{O-RAN Distributed Unit}
\newacronym{TRL}{TRL}{Transformer-RL}
\newacronym{UE}{UE}{User Equipment}

\begin{document}

\title{\vspace{-4mm} Generative AI for Intent-Driven Network Management in 6G RAN: \\ A Case Study on the Mamba Model \\ \vspace{-4mm}}
\vspace{-5pt}

\author{Md~Arafat~Habib, Medhat~Elsayed, Yigit~Ozcan,
        Pedro~Enrique~Iturria-Rivera, Majid~Bavand,\\
        and~Melike~Erol-Kantarci,~\IEEEmembership{Fellow, IEEE}
        \vspace{-20pt}

\thanks{Md~Arafat~Habib, and Melike Erol-Kantarci are with the School of Electrical Engineering and Computer Science, University of Ottawa, Ottawa, Canada (\{mhabi050, melike.erolkantarci\}@uottawa.ca,).}
\thanks{Pedro~Enrique~Iturria-Rivera, Medhat Elsayed, Majid Bavand and Yigit Ozcan are with Ericsson (e-mail:\{pedro.iturria.rivera, medhat.elsayed, majid.bavand, yigit.ozcan\}@ericsson.com).
}
}

\maketitle

\renewcommand{\headrulewidth}{0pt} 

\begin{abstract}
With the emergence of 6G, mobile networks are becoming increasingly heterogeneous and dynamic, necessitating advanced automation for efficient management. Intent-Driven Networks (IDNs) address this by translating high-level intents into optimization policies. Large Language Models (LLMs) can enhance this process by understanding complex human instructions, enabling adaptive and intelligent automation. Given the rapid advancements in Generative AI (GenAI), a comprehensive survey of LLM-based IDN architectures in disaggregated Radio Access Network (RAN) environments is both timely and critical. This article provides such a survey, along with a case study on a selective State-Space Model (SSM)-enabled IDN architecture that integrates GenAI across three key stages: intent processing, intent validation, and intent execution. For the first time in the literature, we propose a hierarchical framework built on Mamba-SSM that introduces GenAI across all stages of the IDN pipeline. We further present a case study demonstrating that the proposed Mamba architecture significantly improves network performance through intelligent automation, surpassing existing IDN approaches. In a multi-cell 5G/6G scenario, the proposed architecture reduces quality of service drift by up to $70\%$, improves throughput by up to $80$ Mbps, and lowers inference time to $60$–$70$ ms, outperforming GenAI, reinforcement learning, and non-machine learning baselines.
\end{abstract}


\vspace{-5pt}
\section{Introduction}

\gls{6G} networks are anticipated to support a diverse set of user requirements and have more complex deployments \cite{1}. Traditional network management methods, which are based on manual configurations and human expertise, will face challenges when optimizing and operating such complex networks. Furthermore, manual configurations are costly, prone to errors and not scalable \cite{1}. \gls{IDN} management is emerging as a solution to the challenges \cite{1}. 

An intent defines a desired network outcome by specifying target metrics without detailing the underlying implementation steps, thereby ensuring both human comprehensibility and machine interpretability \cite{0,3}. Advances in \gls{NLP}, driven primarily by the rise of \glspl{LLM} have enabled operators to express intents more naturally and effectively. Beyond intent processing via \glspl{LLM}, recent works have incorporated other forms of \gls{GenAI} across the \gls{IDN} pipeline, including transformer-based traffic prediction for intent validation \cite{3} and policy execution via Decision Transformers \cite{19}. These \gls{GenAI}-enabled components in \gls{IDN} management collectively reduce manual workload, lower configuration-related errors, and accelerate service deployment and adaptation \cite{4}.

However, an important limitation in existing works is the adoption of transformer-based \gls{GenAI} models at every stage of \gls{IDN}, including intent processing, predictive validation, and policy execution. This introduces a quadratic complexity bottleneck that increases inference latency and limits scalability as sequence lengths grow. To address this issue, a selective \gls{SSM} named Mamba is selected. Mamba provides a suitable alternative by enabling linear-time inference and efficient long-range dependency modeling. 

Mamba extends the \gls{SSM} family by maintaining a recurrent hidden state while replacing fixed linear dynamics with a selective state-update mechanism. This design replaces the Transformer’s quadratic self-attention with a fast and selectively updated memory stream \cite{9}. In \glspl{LLM}, this enables long-context reasoning without computational explosion and supports efficient processing of large text inputs, logs, and multi-page instructions. In decision-making settings, such as Decision Mamba for \gls{RL} \cite{16}, selective memory retains only historically meaningful transitions and avoids uniform weighting of past states. In time-series forecasting, Mamba captures both smooth temporal trends and sudden spikes without relying on positional encodings that limit extrapolation \cite{20}. Across these domains (\gls{LLM}, time-series forecasting, and control decision making), Mamba addresses the structural limitations of Transformers while offering a faster, lighter, and more stable alternative.

Motivated by the need for more efficient \gls{GenAI} techniques in hierarchical and disaggregated \gls{RAN} environments, this work introduces Mamba as an alternative to transformer-based algorithms for \gls{IDN} management. We begin with a brief review of state-of-the-art \gls{GenAI} methods used across the \gls{IDN} workflow. We then propose a three-stage Mamba-driven methodology that covers intent processing, validation, and execution. Building on this methodology, we also show how each stage can be realized within a hierarchical \gls{RAN} architecture through multiple specialized Mamba-based models. A case study further demonstrates the practical realization of the entire pipeline. Specifically, intents are interpreted using a parameter-efficient Mamba-based \gls{LLM} fine-tuned on a custom dataset \cite{21}. Intent feasibility is assessed using a Mamba-based time-series predictor that forecasts key network parameters and validates operator objectives accordingly. Finally, intent execution is performed via Decision Mamba, for its superior memory and computational efficiency compared to transformer-based approaches \cite{9}.  

The core contributions of this paper are threefold. First, it presents a precise, yet comprehensive survey of \gls{GenAI}-based \gls{IDN} architectures, highlighting their main features, limitations and applications. Along with the survey, we also present open research directions, practical deployment considerations, and standardization implications for \gls{GenAI}-enabled \gls{IDN} management, offering a roadmap for future development in this domain. Second, we introduce the first Mamba-driven \gls{IDN} framework in the literature and demonstrate how it can replace traditional transformer-based components to deliver linear-time inference with improved long-range reasoning. Third, we show the effectiveness of the proposed framework through a detailed case study within a hierarchical \gls{RAN} architecture that showcases the end-to-end workflow of Mamba-based \gls{IDN} management and highlights its computational and operational advantages.

Unlike prior \gls{GenAI}-enabled \gls{IDN} frameworks that rely on transformer-based models and typically introduce intelligence at only one or two stages of the \gls{IDN} pipeline, this work is the first to systematically replace transformers across intent processing, validation, and execution using the Mamba models. This unified Mamba-driven architecture eliminates quadratic attention overhead, enables faster inference, and improves crucial \glspl{KPI}.

In our evaluation across a multi-cell \gls{5G}/6G scenario, the proposed Mamba–enabled \gls{IDN} architecture delivers substantial performance gains across all major \glspl{KPI}. The framework consistently achieves the lowest \gls{QoS} drift, reducing intent-violation rates by up to $70\%$ compared to the baselines. It further enhances network throughput by $45$--$80$ Mbps under moderate to high traffic loads and achieves up to two times higher energy efficiency relative to the baselines. In addition, the proposed Mamba architecture for \gls{IDN} management enables fast inference, lowering decision latency to $60$--$70$ ms, which is considerably better than the baselines as well.

\section{\gls{GenAI}-Based Intent-driven Network Management Schemes}

\label{s2}

\glspl{LLM} have transformed \gls{IDN} management due to their ability to perform zero-shot, few-shot, and fine-tuned learning. They now represent the state-of-the-art for intent recognition and processing, and most existing \gls{IDN} studies employ them to interpret operator intents. The literature has also evolved beyond intent processing, incorporating AI-based modules for validating and executing intents \cite{1,3}. Recently, a full-scale \gls{GenAI}-driven solution was proposed in \cite{19}. Motivated by these developments, this section first reviews works that focus specifically on \gls{GenAI}-based methods directly supporting the \gls{IDN} management workflow across intent processing, validation, and execution. We summarize reviewed works in terms of their main features, challenges, and applications in Table \ref{tab1}. Note that we have selected works that have used GenAI in at least one of the stages of the \gls{IDN} management workflow. Subsequently, in this section, we explore the open research directions, 3GPP and Open \gls{RAN} standard implications, and practical deployment considerations for \gls{GenAI}-based \gls{IDN} management. 

\begin{table*}[!t]
    \caption{\gls{LLM}-Based \gls{IDN} Management Approaches: Features, Challenges, and Applications}
    \centering
    \renewcommand{\arraystretch}{1.3}
    \begin{tabular}{|m{4.5cm}|m{4.5cm}|m{3.6cm}|m{3.6cm}|} 
        \hline
        \multicolumn{1}{|c|}{\textbf{Category}} & 
        \multicolumn{1}{c|}{\textbf{Main Features}} & 
        \multicolumn{1}{c|}{\textbf{Challenges}} & 
        \multicolumn{1}{c|}{\textbf{Applications}} \\
        \hline
        \gls{LLM}-assisted intent processing and management for \gls{5G} core networks \cite{11,12}
        & \begin{itemize}
            \item Supports multiple intent transmission to the model.
            \item Enhances network orchestration with semantic routing.
            \item Prevents \gls{LLM} hallucination. 
            \item Includes a \gls{RAG} module. 
          \end{itemize}
        & \begin{itemize}
            \item Reliance on \glspl{LLM}, which are not inherently designed for network control and automation. 
            \item Distinguishing between similar or overlapping intents.
          \end{itemize}
        & \begin{itemize}
            \item Conversion of operator-defined intents into slice deployment policies. 
            \item Mapping intents to edge-based service deployments.
          \end{itemize} \\
        \hline
        
        \gls{LLM}-Driven multi-agent and negotiation-based intent management for \gls{6G} networks \cite{13,4}
        & \begin{itemize}
            \item Employs \gls{MARL} and \gls{LLM}-driven negotiation.
            \item Resolves conflicting intents in multi-stakeholder settings.
          \end{itemize}
        & \begin{itemize}
            \item Cross-domain conflict resolution.
            \item Fairness in resource allocation.
            \item Scalability concerns.
          \end{itemize}
        & \begin{itemize}
            \item \gls{6G} network slicing.
            \item Dynamic spectrum allocation.
            \item Autonomous resource management.
          \end{itemize} \\
        \hline

        End-to-end AI-based intent-driven network automation \cite{1,3,19,22}
        & \begin{itemize}
            \item Converts high-level intents into actionable network policies.
            \item Verifies if an intent is feasible given the network’s current or predicted state.
            \item End-to-end automation using AI.
          \end{itemize}
        & \begin{itemize}
            \item Computational complexity of processing and validating multiple intents in real-time increases exponentially.
          \end{itemize}
        & \begin{itemize}
            \item Zero-touch network configuration.
            \item Predictive maintenance and fault recovery.
          \end{itemize} \\
        \hline

        Generative AI and multimodal intent-based network management \cite{15}
        & \begin{itemize}
            \item Uses multimodal generative AI for intent translation.
            \item LLMs are tuned with few-shot learning.
          \end{itemize}
        & \begin{itemize}
            \item High dependency on predefined templates.
            \item Expansion to include multimodal inputs.
          \end{itemize}
        & \begin{itemize}
            \item Automation of the deployment and management of network slices.
          \end{itemize} \\
        \hline
    \end{tabular}
    \label{tab1}
\end{table*}

\subsection {State-of-the-art \gls{GenAI}-Based Approaches for \gls{IDN} Management}

\subsubsection{\gls{LLM}-Assisted Intent Processing and Management for \gls{5G} Core Networks}

Two recent studies explore \gls{LLM}-assisted approaches on intent-based management in \gls{5G} core networks. Manias et al. in their work \cite{11} explore \gls{LLM}-driven intent extraction as a key enabler of next-generation zero-touch network service management. A customized \gls{LLM} model is designed to interpret and translate user intents into actionable network policies. By utilizing \glspl{LLM}, the framework reduces human intervention while maintaining adaptability across diverse network management tasks.

Semantic routing is introduced to refine \gls{LLM}-assisted intent-based networking in \cite{12}. Traditional \gls{LLM}-driven approaches face issues such as hallucination, scalability limitations, and accuracy degradation when handling complex network intents. To address these challenges, the proposed framework integrates a semantic router, which ensures deterministic decision-making by routing extracted intents through predefined pathways rather than relying solely on \gls{LLM}-generated responses. 

\subsubsection{\gls{LLM}-Driven Multi-Agent and Negotiation-Based Intent Management}

The framework proposed in \cite{13} is a collaborative, multi-agent system for managing shared network resources in \gls{6G}. This system deploys \gls{LLM}-based agents to represent different business entities. Each of these entities negotiates service-level objectives such as throughput, cost efficiency, and energy savings. The framework acts as a central mediator to resolve conflicts by utilizing \glspl{LLM} alongside optimization techniques and real-time network observability.

Another study \cite{4} introduces a comprehensive \gls{LLM}-driven intent life cycle management system designed to handle all stages of intent processing, from decomposition and translation to negotiation, activation, and assurance. 

\subsubsection{End-to-End AI-Based Intent-Driven Network Automation}

Recent studies introduce end-to-end AI frameworks that integrate \glspl{LLM}, algorithms like \gls{DRL}, \gls{MARL}, and \gls{HRL} to realize fully automated network management. 

A recent study introduces a transformer-based time series predictor to validate intents before execution \cite{3}. This predictive validation module analyzes historical network data and forecasts traffic patterns to ensure that requested optimizations do not negatively affect service quality. Once an intent is validated, an \gls{HRL} framework selects and triggers appropriate network optimization applications such as beamforming, traffic steering, and power control. An attention-based \gls{HRL} model filters out suboptimal actions to reduce computational overhead while maximizing efficiency.

Zhang et al. propose an intent-driven closed-loop control and management framework for \gls{6G} Open \gls{RAN}, which integrates AI across all stages of the \gls{IDN} pipeline \cite{22}. Intent translation is performed using a long short-term memory model with morphological rules, while intent verification and decomposition rely on AI-assisted reasoning via event calculus. The refined objectives feed into a control loop, where a \gls{DQN}-based network application (near-real-time control) executes slice-level resource allocation. 

Another recent contribution is proposed in \cite{19}, which demonstrates a fully \gls{GenAI}-enabled workflow for intent processing, validation, and execution. The framework integrates a \gls{QLoRA}-fine-tuned \gls{LLM} for intent interpretation, an Informer model for long-horizon traffic prediction, and a goal-aware \gls{DT} for RAN control. By combining large-scale language understanding with predictive analytics and sequential decision-making, this work illustrates how \gls{GenAI} can support seamless end-to-end automation in \gls{6G} Open \gls{RAN} environments.

Another breakthrough in end-to-end AI-driven network automation is the integration of multi-agent learning frameworks \cite{1}, where AI-driven network agents negotiate and resolve conflicting intents in real-time. This is particularly essential in multi-tenant \gls{6G} networks, where various stakeholders have competing resource demands. 

\subsubsection{Generative AI and Multimodal Intent-Based Network Management}

A multimodal intent recognition framework presented in \cite{15} utilizes \glspl{LLM} with \gls{GenAI} to process diverse inputs like text, images, and deployment descriptors to enable network operators for specifying intents through text prompts, graphical designs, or configuration files. AI dynamically refines intents to ensure accuracy while aligning with service-level objectives.

A key application of this approach is network-as-a-service orchestration, where \gls{LLM}-powered agents automate network slice management. The framework enables zero-touch deployment to map business requirements to network slice templates.

\subsection{Open Research Directions for Enhancing \gls{IDN} Management}

The key limitations of the \gls{GenAI}-based \gls{IDN} management schemes discussed so far in this section can be described as follows: \gls{LLM} hallucination and interpretability challenges, which can result in inaccurate or overly generalized outputs; absence of a real-time feedback mechanism, making it difficult to correct errors in network policies after deployment; and deployment challenges, particularly due to memory constraints and high computational requirements. To address these issues, we recommend some possible measures as open research directions.

$\bullet$~\gls{RAG}: We can use network-specific databases and retrieval models to ensure \glspl{LLM} only generate responses based on verified knowledge to reduce hallucination. Particularly, there has not been any work in the \gls{IDN} literature that has explored Agentic RAG (using LLM agent to choose between databases).

$\bullet$~\gls{RLHF}: Since re-training is not always an option due to computational resource scarcity, \gls{RLHF} can be a powerful technique for improving \gls{LLM}-based \gls{IDN} management to address the issue of real-time adaptation and feedback loops. We can define a reward function based on network \glspl{KPI} (e.g., latency, throughput, energy efficiency). Then, a feedback loop can be implemented where operators and simulated environments score the \gls{LLM}’s intent outputs. Lastly, \gls{RL} algorithms can be used to refine \gls{LLM}-generated responses.

$\bullet$~\gls{LLM} quantization and compression: Computation overhead caused by \gls{LLM} deployments can be handled by implementing 4-bit quantization techniques (\gls{QLoRA}, \gls{GPTQ}) to reduce \gls{LLM} model size while maintaining high accuracy.

$\bullet$~Mamba models: Mamba models can be incorporated for reduced memory overhead and faster inference instead of the transformers in the \gls{IDN} management pipeline.

\subsection{Practical deployment considerations}

Practical deployment of \gls{GenAI}-enabled \gls{IDN} management must account for the strict latency requirements of \gls{RAN} environments, where many control-loop decisions must be completed within the near-real-time window of $10$ms to $1$s. These constraints limit the size and complexity of deployable \glspl{LLM} and predictive models. Another key challenge is ensuring robustness against \gls{QoS} and intent drift, as real-world network conditions and operator objectives may shift over time. To maintain reliable operation, deployment pipelines must incorporate lightweight model updates, drift detection mechanisms, and safeguards that ensure \gls{GenAI}-driven decisions remain compliant with \gls{QoS} guarantees. 

\subsection{3GPP and Open RAN standard implications}

\gls{GenAI}-enabled \gls{IDN} management has several implications for ongoing 3GPP and Open \gls{RAN} standardization efforts. First, 3GPP Release 17 and beyond define intent-driven management services (TS 28.312) \cite{0}, but current specifications assume rule-based intent translation. Integrating \gls{LLM}-based intent processing and \gls{SSM}-driven validation requires extending these standards to support natural-language intent formats, semantic models, and AI-assisted intent interpretation pipelines. Second, predictive validation and goal-aware decision-making must align with 3GPP’s management data models, where AI-generated recommendations map to standardized network resource models and service exposure \glspl{API}.

From the Open \gls{RAN} perspective, introducing \gls{GenAI} into the intent execution loop affects the interfaces between the service management orchestrator, non-real-time intelligent controller, near-real-time intelligent controller, and network applications inhabiting in those controllers. Mamba-based predictors and decision models must comply with O-RAN open interfaces such as A1 (connects the non and near-real-time controllers), E2 (links the near-real-time controller with the distributed RAN nodes (\gls{O-DU}/\gls{O-CU}) for cross-layer orchestration. Moreover, near-real-time latency requirements ($10$ms–$1$s) impose constraints on the deployment of \gls{GenAI} models that motivate faster inference and hardware-aware optimization. 

\section{Three-step Methodology for \gls{IDN} Management and its \gls{RAN} Integration with \gls{GenAI}}

\label{s3}

In this section of the paper, we discuss a three-step framework consisting of intent processing, validation and execution to introduce \gls{IDN} management and how \gls{GenAI} can be integrated within it. This section will help the readers have the necessary background to understand the Mamba-based case study to be presented in the next section.   

\begin{figure}[!t]
\centerline{\includegraphics[width=0.7\linewidth]{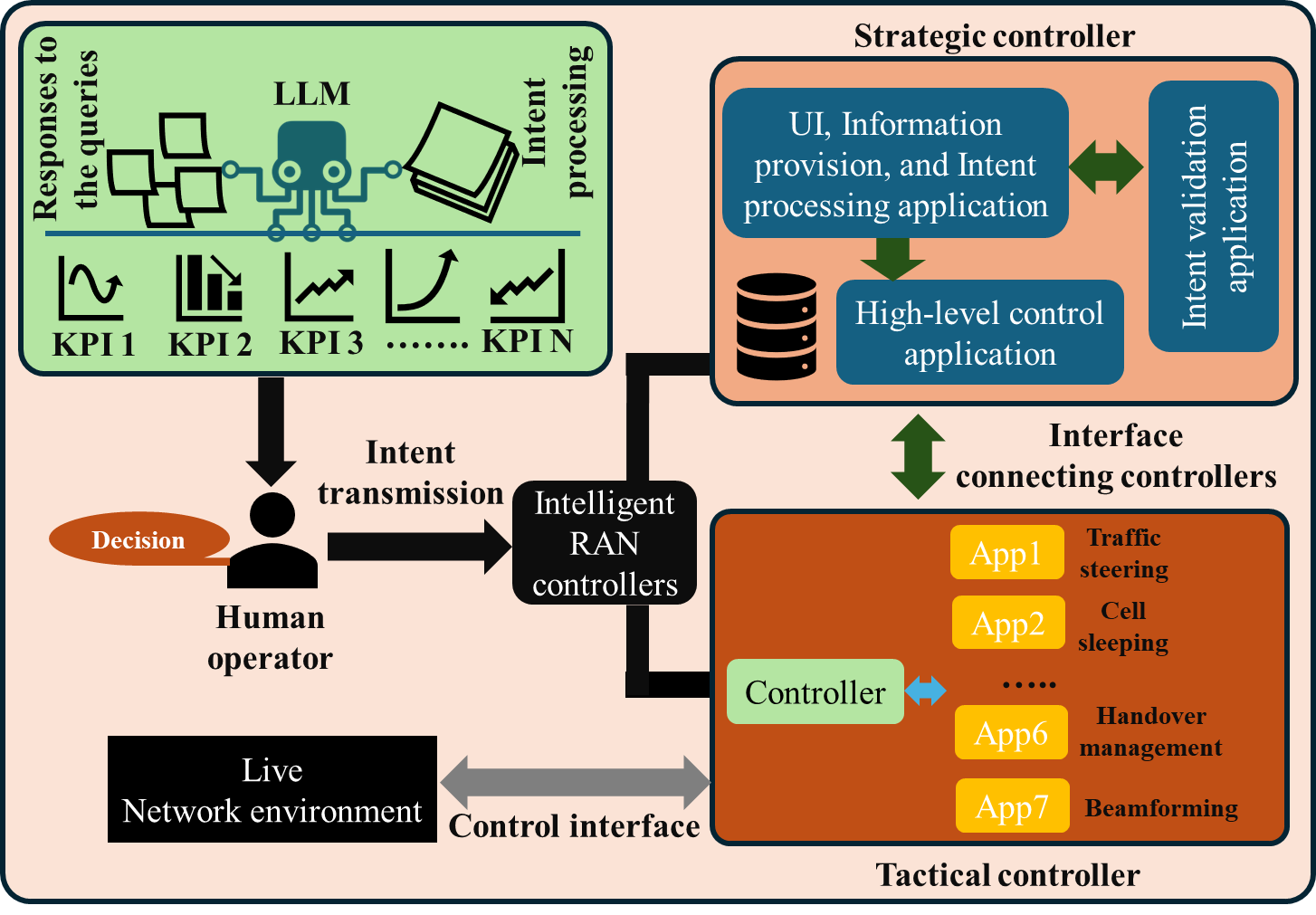}}
\caption{Integration of the proposed \gls{IDN} management framework with hierarchical \gls{RAN} architecture.} 
\vspace{-10pt}
\label{fig2}
\end{figure}

$\bullet$~Intent processing: Intent processing is the phase where high-level, human-expressed network intents (e.g., ``Minimize latency for video traffic'') are interpreted, extracted, and translated into structured, machine-executable commands using \glspl{LLM}, multimodal AI, or rule-based systems. It is the first step of the \gls{IDN} management and is highly important because if intent interpretation is not done correctly, it can lead to performance degradation.

$\bullet$~Intent validation: Intent validation is the process of assessing whether the current network conditions and capacity can accommodate a given intent or a set of intents. This involves verifying whether the available resources, traffic conditions, and operational constraints allow for the successful execution of the desired network optimization actions. Intent validation ensures that a processed intent is feasible, conflict-free, and optimal before execution.

$\bullet$~Intent execution: Intent execution is the final stage of \gls{IDN} management, where validated intents are transformed into real-time network actions. This phase involves invoking appropriate network optimization policies to dynamically configure resources and ensure that the desired service-level objectives are met efficiently. Intent execution can be realized through network function orchestration, AI-driven decision-making, or \gls{RL}-based adaptation.

\begin{figure*}[!t]
\centerline{\includegraphics[width=0.7\linewidth]{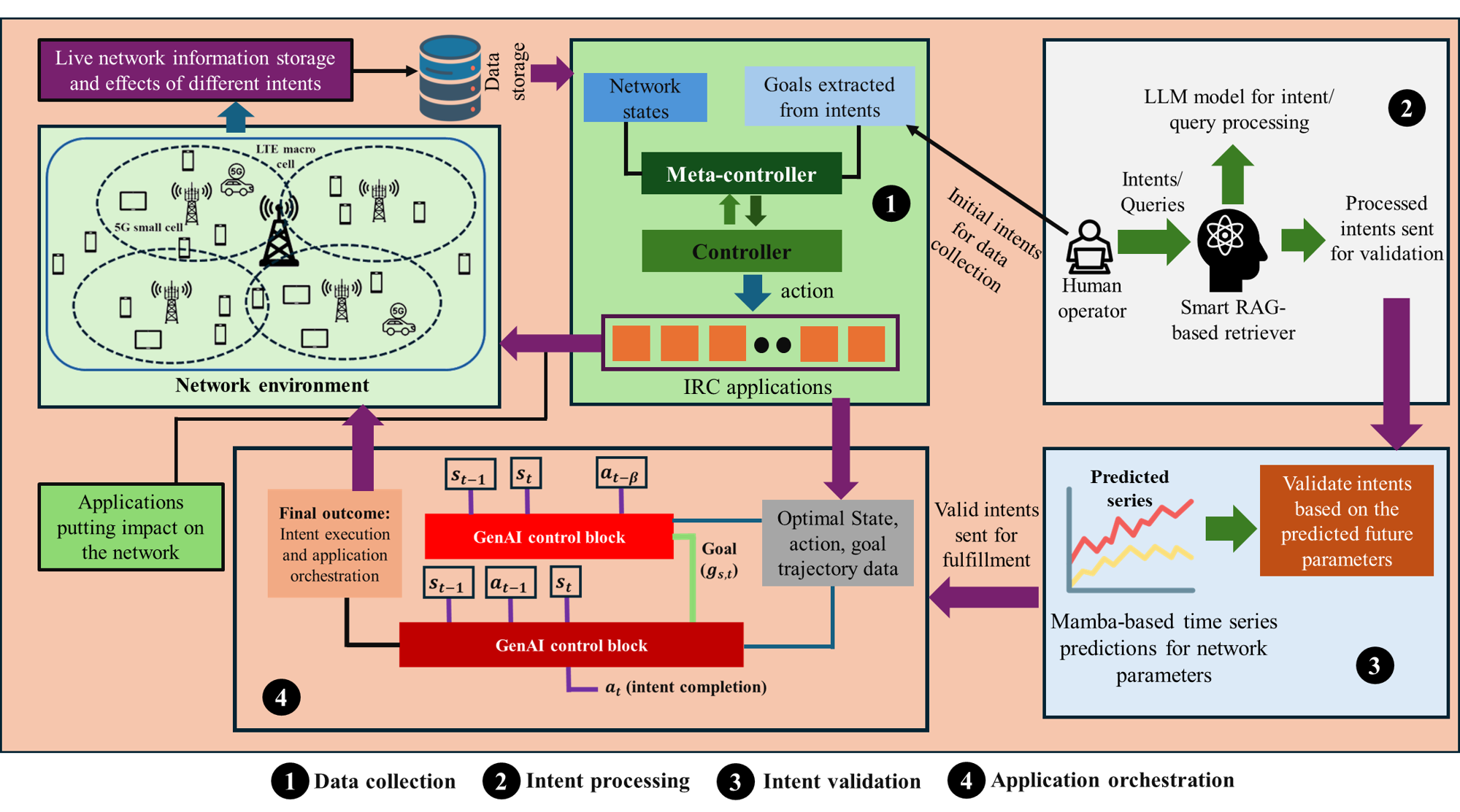}}
\caption{End-to-end GenAI framework for intent-driven network management.} 
\vspace{-10pt}
\label{fig3}
\end{figure*}

\subsection{Hierarchical \gls{RAN} Integration with the three-step \gls{IDN} Management Architecture}

Modern hierarchical \gls{RAN} architectures, such as Open \gls{RAN}, use a multi-layered and modular design to support flexibility, scalability, and interoperability in next-generation wireless networks. These systems break down the traditional monolithic \gls{RAN} into software-driven components that enable automation and AI-based optimization. In this work, we adopt a two-level architecture composed of a strategic controller and a tactical controller. The strategic controller manages long-term objectives like service quality, energy efficiency, and traffic patterns, while the tactical controller focuses on real-time operations, including handovers, resource allocation, and traffic steering. This layered approach helps implement intent processing, validation, and execution by supporting structured decisions across different time scales and abstraction levels.

We provide an example of implementing the three-step \gls{IDN} management workflow using a custom hierarchical \gls{RAN} architecture in Fig. \ref{fig2}. A human operator can observe the current network status or query for any relevant network information before providing an intent. The \gls{LLM} module in the picture is given as an example. The operator may use logs or another query-based database to gather information instead of using an \gls{LLM}. After that, the operator can provide an intent that is received by the user interface application in the top-level \gls{IRC} named strategic controller. It then passes through the intent validation application. This application can be designed to perform intent validation via rule-based static methodologies or \gls{GenAI}-based predictive analysis. Lastly, at the bottom, there is a tactical controller responsible for intent execution. In Fig. \ref{fig2}, we provide an example of intent execution via network applications in the tactical controller. Based on the intent, multiple network applications such as traffic steering, power allocation, and cell sleeping can be initiated and orchestrated to fulfill an intent. 

\section{A Case Study on Mamba Model for \gls{IDN} Management}
\label{s4}

Hierarchical use of \gls{GenAI} algorithms can simplify complex decision-making by splitting learning across multiple layers. At the top layer, \glspl{LLM} interpret human intents and extract key metrics and constraints. These insights guide a mid-tier time-series predictor that analyzes historical and real-time data to validate network conditions. At the lowest layer, control algorithms translate validated intents into executable policies. This layered design improves efficiency, lowers computational overhead, and supports adaptive, real-time decision-making.

Fig. \ref{fig3} shows the end-to-end \gls{GenAI} framework for \gls{IDN} management used in our case study. Data collection follows an \gls{HRL} structure with two decision-making levels. The higher-level controller sets goals based on operator intents, such as raising throughput or lowering power use. The lower-level controller chooses the appropriate network application based on traffic load, energy efficiency, or packet loss. Five applications are used: traffic steering, cell sleeping, beamforming, power allocation, and energy-efficient handover. Each is implemented using \gls{DRL}. Traffic steering distributes traffic across different radio access technologies to improve delay and throughput. Cell sleeping reduces power by disabling underutilized base stations. Beamforming computes optimal angles using user location, while power allocation improves total throughput through smart resource assignment. The handover management application enhances energy efficiency through adaptive handover decisions. After selection, each application independently adjusts its parameters using its own policy. More details are available in \cite{3}.

After each action, the system evaluates its impact by checking whether it improves the targeted network metrics. A reward mechanism measures how effective the selected applications were in enhancing performance. Over time, this process produces a large dataset containing past actions, network states, and observed outcomes. The dataset is generated using a custom simulator designed to mimic realistic network conditions and can later be enriched with real-world traces to further improve accuracy.

The collected data is organized into three datasets. The first supports \gls{LLM} fine-tuning and contains query–response pairs and intent-driven prompts to help the model interpret operator goals. The second dataset contains time-series measurements of parameters such as traffic load, packet loss, and power consumption. This data enables predictive intent validation by forecasting whether an intent is safe to execute. The third dataset is a trajectory set that trains the control algorithms. It records past decisions, states, and outcomes, enabling decision models to learn effective application selection that meets operator objectives. 

In this case study, we adopt Mamba-\gls{PEFT} to fine-tune a pretrained Mamba-based \gls{LLM} for intent-driven \gls{RAN} management \cite{21}. Instead of updating all model parameters, which would require high memory and computing, Mamba-\gls{PEFT} modifies only selected components inside the Mamba block. These include the input and output projections, specific state-space transitions, or a small number of additional state dimensions. By limiting the number of trainable parameters, the method reduces memory usage and keeps the fine-tuning process efficient while preserving the model’s long-context reasoning ability. Since Mamba processes sequences in linear time, the overall adaptation becomes more scalable than transformer-based alternatives. The fine-tuned model is integrated with a \gls{RAG} module, which provides access to current network information, configurations, and logs. With both selective parameter adaptation and retrieval-augmented knowledge, the resulting Mamba-\gls{PEFT} model delivers accurate and context-aware responses to operator queries while remaining suitable for use in hierarchical \gls{RAN} environments.

In the intent validation stage, the system evaluates whether an operator’s intent can be executed safely under both current and predicted network conditions. \glspl{KPI} such as traffic load, packet loss, and power consumption are continuously monitored and forecasted using Mamba4Cast \cite{20}. Mamba4Cast is a Mamba-based zero-shot time-series foundation model built on selective state-space representations. It combines causal convolution layers with lightweight Mamba-based encoder blocks to capture multi-scale temporal patterns and generates multi-horizon forecasts in a single pass. When an operator submits an intent to increase throughput or reduce power usage, the predicted \gls{KPI} trajectories are checked against predefined feasibility thresholds. If the forecasts indicate a risk of performance degradation, such as requesting higher throughput during a period of high expected load, the system marks the intent as infeasible. A lookup table that maps predicted KPI patterns to known quality-of-service outcomes from past executions further accelerates the decision. Through this combination of Mamba-based time-series forecasting and historical pattern matching, the validator ensures that only intents suitable for future conditions proceed to execution.

Finally, the last part of this case study includes the Mamba architecture for intent execution via network application orchestration. In particular, we use the Decision Mamba proposed in \cite{16}. We propose \gls{HDMGA}, a hybrid hierarchical decision-making framework that integrates Decision Mamba as the high-level decision mechanism and a \gls{DT} as the low-level control transformer.

At a high level, Decision Mamba serves as the goal-aware memory mechanism, responsible for retaining and retrieving critical past actions that have significantly contributed to achieving network optimization objectives. Unlike conventional self-attention mechanisms, which perform exhaustive searches over past trajectories, Mamba employs a selective \gls{SSM} with a learnable dynamic memory retention mechanism. This allows the system to selectively remember only the most relevant past actions while discarding irrelevant ones. As a result, computational overhead gets reduced, and decision efficiency improves. 

At the low level, the \gls{DT} \cite{19} functions as the control transformer, refining real-time actions based on the goal and past knowledge retrieved by the Decision Mamba. Once Mamba identifies the most impactful past action, it is passed to the control transformer, which uses a sequence modeling approach to predict the optimal action for the current state. By conditioning decisions on goal awareness rather than predefined reward sequences, the control transformer ensures that each selected action aligns with the operator’s high-level intent.

The intent format, inputs, and outputs used across the pipeline are as follows. For intent processing, the input is either a natural-language query requesting contextual network information or an optimization intent specifying a \gls{KPI} and a desired magnitude. The output is either an \gls{LLM}-generated informational response or a structured representation identifying the target \gls{KPI} and magnitude, respectively. For intent validation, the input is the parsed optimization intent together with predicted network \glspl{KPI} generated by the forecasting module, and the output is a feasibility decision indicating whether the intent can be safely executed under current or predicted conditions. For intent execution, the input is the desired magnitude of improvement of a \gls{KPI}, and the output is a set of executable network actions, including the selected applications and their control parameters, deployed in the \gls{RAN} controllers.

In this case study, when the operator provides an intent such as ``Increase throughput by 10\%'', the system sets this increased demand for throughput as its goal. It then recalls a past scenario with similar network conditions, such as high traffic load and moderate packet loss, where a specific combination of applications improved throughput. If the current network state does not support the intent, the system activates a feedback loop that either prompts the operator to revise the intent or uses the \gls{LLM} to suggest feasible alternatives. The top-level \gls{GenAI} block (Decision Mamba) reviews the goal and the recent network context, retrieves the successful past action, and uses it to guide the present decision. The control transformer described before has the current state of the network and the past action suggested by the top-level decision Mamba block. It uses this context to decide what to do next. It may decide to take the action of enabling a traffic steering application along with a power control application. This decision is tailored to current needs while being inspired by past success, aiming to fulfill the operator’s intent efficiently.

\begin{figure}[!t]
\centerline{\includegraphics[width=0.95\linewidth]{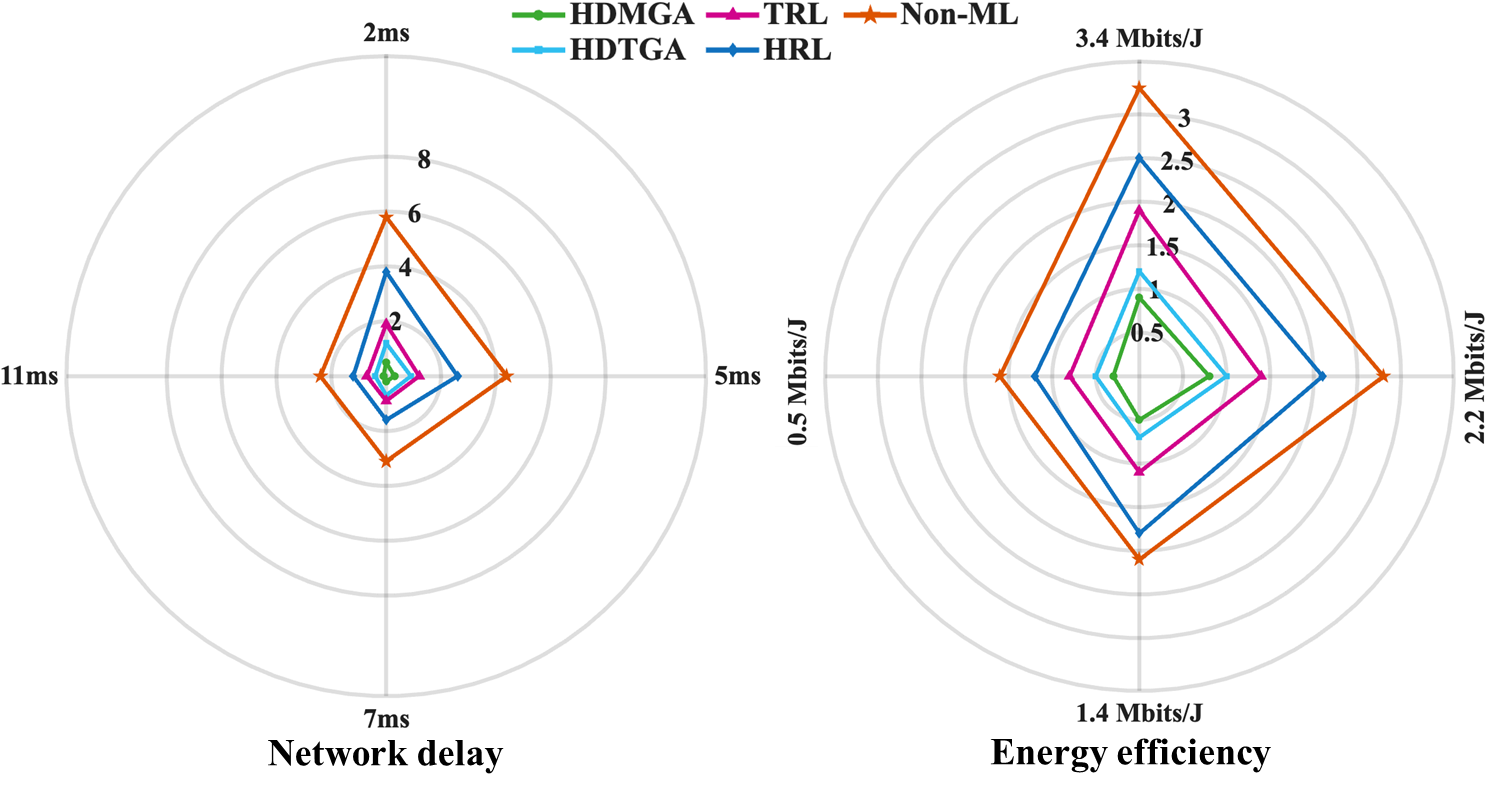}}
\caption{Performance comparison of the proposed method against the baselines in terms of goal deviation (network delay and energy efficiency).}
\vspace{-10pt}
\label{fig4}
\end{figure}

\begin{figure}[!t]
\centerline{\includegraphics[width=0.95\linewidth]{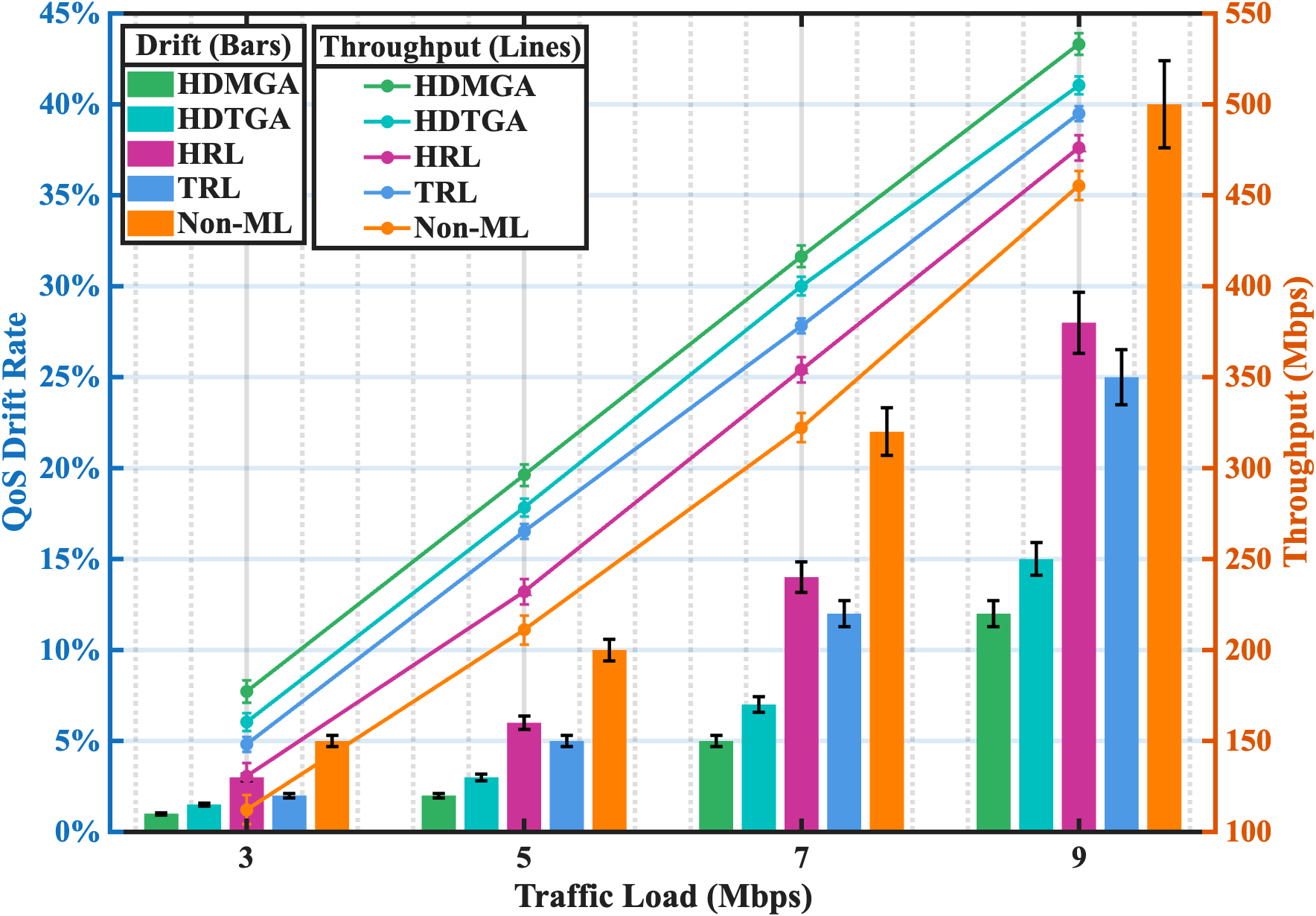}}
\caption{Performance comparison of the proposed method against the baselines in terms of throughput and reliability (\gls{QoS} drift).}
\vspace{-20pt}
\label{fig6}
\end{figure}

To evaluate the effectiveness of \gls{HDMGA}, we compare it against four baseline approaches. First of all, we adopt the deterministic rule-based method, which performs threshold-driven application triggering without any learning capability. This is referred to as Non-ML baseline in our performance graphs. Furthermore, we consider another baseline named \textit{StARformer} where a transformer model is merged into an RL agent that predicts actions directly from recent state–action–reward tuples \cite{23}. We address this as \gls{TRL} in our performance graphs. The last two baselines are \gls{HDTGA} and \gls{HRL} with intent validation. The \gls{HDTGA} framework, which is our prior method, utilizes a hierarchical decision transformer architecture. A meta-transformer searches past trajectories to retrieve the most relevant past action, which is then used to guide the control transformer in predicting future actions. The \gls{HRL} with intent validation baseline follows a conventional \gls{HRL} approach with a dedicated intent validation mechanism \cite{3}. 

The simulation setup in this case study consists of a macro cell surrounded by densely deployed small cells in a multi-\gls{RAT} environment, serving $60$ users. The \gls{5G} \gls{NR} operates at $3.5$ GHz (mid-band) and $30$ GHz (high-band) with bandwidths of $50$ MHz and $100$ MHz, respectively. The max transmission power of the 5G NR BS is $43$ dBm. On the other hand, \gls{LTE} operates at $800$ MHz with a $20$ MHz bandwidth and $38$ dBm maximum transmission power.

The simulator models the behavior of the \gls{RAN} using a 3GPP-aligned design that includes realistic traffic patterns, mobility, and Radio Frequency (RF) impairments. Four traffic types are generated each having its own inter-arrival time and distribution, which creates a mix of bursty, periodic, and low-latency traffic. Traffic types include video, gaming, voice, and vehicle-to-base station data traffic (\gls{URLLC} use case), with packet inter-arrival times of $12.5$ ms, $40$ ms, $20$ ms, and $0.5$ ms, respectively, following Pareto, Uniform, and Poisson distributions. User mobility follows 3GPP system-level guidelines, where \gls{UE} positions are updated every \gls{TTI} and movement directly affects path loss, Doppler shift, beam directions, and handover events. The RF environment includes \gls{CDL}-based multipath fading, shadowing, thermal noise, beam misalignment, and multi-cell interference. These impairments evolve with mobility and produce realistic fluctuations in signal-to-interference-plus-noise ratio, throughput, and delay. Together, the traffic models, mobility behavior, and RF effects create a reliable and reproducible simulation environment for evaluating intent-driven RAN management.

Our proof-of-concept is implemented using a multi-\gls{RAT} MATLAB simulation (5G, \gls{LTE} and \gls{CDL} fading) integrated with Python-based Decision Mamba and Mamba-LLM modules. The \gls{LLM} is fine-tuned using a $1.1$k-sample dataset \cite{19} of dynamic network queries, semi-static operational knowledge, and structured intents, while a Mamba-based predictor validates intent feasibility by forecasting load, power, and packet-drop trends.

As presented in Fig. \ref{fig4}, it can be observed that across both delay and energy-efficiency targets, the proposed \gls{HDMGA} achieves the smallest goal deviations among all compared methods. This improvement is due to \gls{HDMGA}’s selective state-space modeling and goal-conditioned hierarchical design, which allow the agent to retain only the most relevant historical transitions and choose application combinations that directly contribute to meeting the operator’s intent. \gls{HDTGA} performs moderately well but still shows higher deviations because its transformer-based attention weighs all past states uniformly and introduces unnecessary computational overhead. \gls{TRL} experiences larger deviations since it lacks hierarchy, sub-goal decomposition, and selective memory, making it less effective when coordinating multiple network applications. \gls{HRL} shows even higher deviations because it must explore the entire combinatorial action space without attention-based filtering or intent validation, which often results in suboptimal application choices under dynamic traffic conditions. The Non-ML heuristic baseline records the largest deviations for both metrics because its static threshold-based rules cannot adapt to changing network states or operator intent magnitudes. While only delay and energy-efficiency figures are presented, the same performance trend is consistently observed for throughput goal deviations as well.

The \gls{QoS} drift rate represents how much the network’s performance deviates from the level expected by the operator when an intent is executed. The percentage is computed by averaging the deviation over a monitoring window that spans several \glspl{TTI}. Fig. \ref{fig6} compares the drift rates of all methods across different traffic loads. The proposed \gls{HDMGA} maintains the lowest drift rates at every load level, indicating that it consistently meets the operator’s expected performance targets with minimal deviation. This reliability comes from its selective state-space modeling and goal-aware hierarchical design, which allow the agent to prioritize only the most relevant past transitions and choose the correct combination of applications even when the network becomes highly loaded. 

The throughput results in Fig. \ref{fig6} follow the same overall trend observed in the drift analysis. As the traffic load increases, \gls{HDMGA} consistently achieves the highest throughput among all methods, reflecting its ability to select the most effective combination of \gls{RAN} applications while respecting operator intent. Its selective memory mechanism helps preserve the most relevant context for scheduling and resource allocation, which leads to more efficient utilization of network capacity.

\begin{figure}[!t]
\centerline{\includegraphics[width=0.95\linewidth]{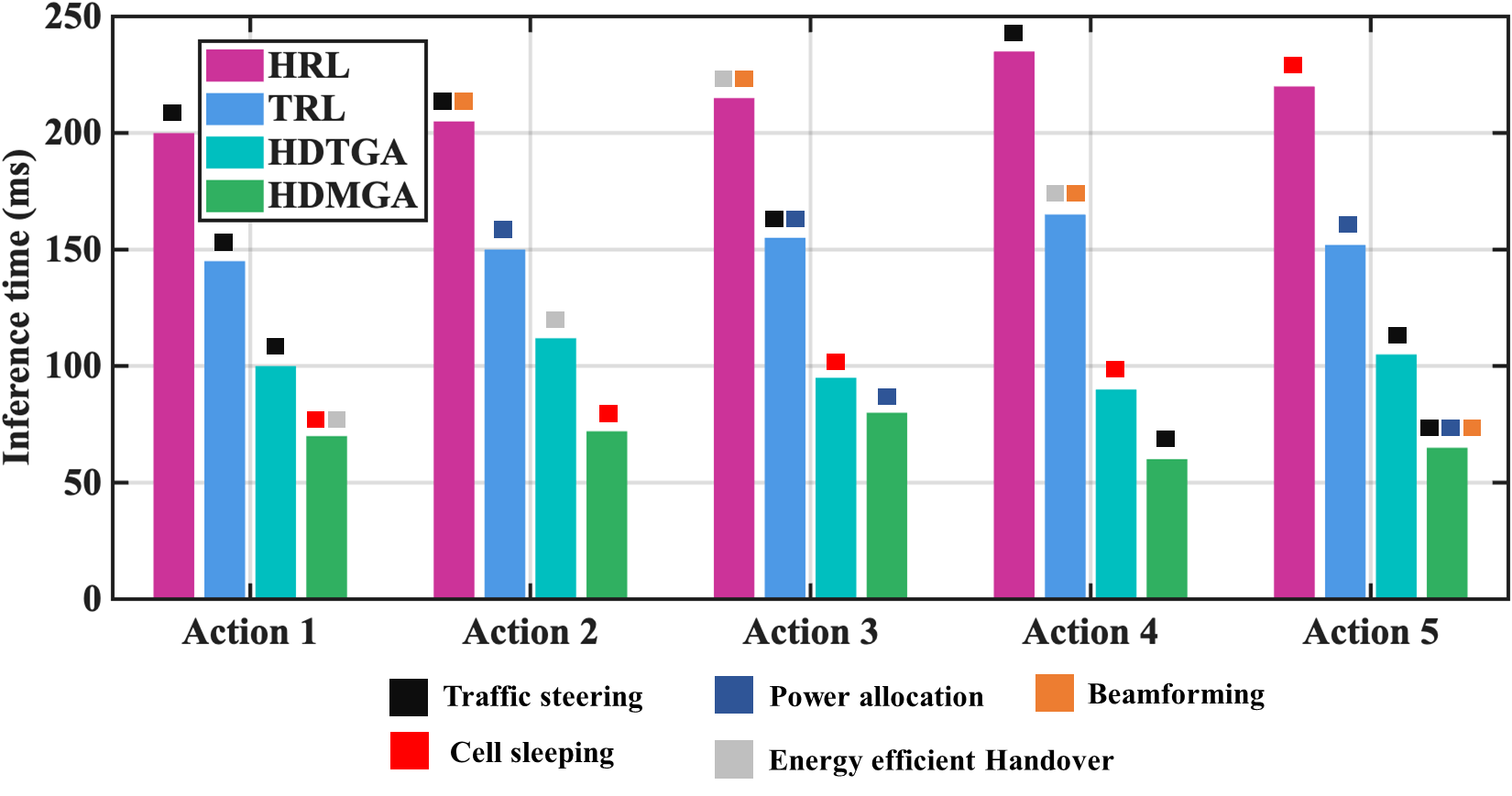}}
\caption{Performance comparison in terms of action inference time.} 
\vspace{-20pt}
\label{fig5}
\end{figure}

Fig.~\ref{fig5} presents a plot showing that the action inference time of the proposed method is lower than that of the baselines. The actions illustrated are not sequential decisions taken under identical network conditions. Instead, each action corresponds to an independent inference made under distinct network states at specific time points during the simulation. Since network conditions (e.g., traffic load, power consumption, packet loss) and operator intents vary over time, different methods naturally select varying combinations of applications based on context. This variability reflects each method's adaptability to dynamic \gls{RAN} environments. The figure demonstrates that \gls{HDMGA} achieves faster decision-making, which is essential for near-real-time network automation. This efficiency is enabled by its memory-guided design, where Decision Mamba selectively retains and retrieves only the most relevant past actions. All the methods in the figure maintain inference times within the near-real-time latency bounds defined by O-RAN specifications ($10$ms to $1$s).

\section{Conclusions}

This article presented a \gls{GenAI}-driven framework for \gls{IDN} management that integrates selective state-space modeling across intent processing, validation, and execution for the first time in the literature. By utilizing a fine-tuned Mamba-based \gls{LLM} for intent interpretation, a Mamba-enabled predictor for proactive validation, and a Mamba-enhanced intent execution module, the proposed architecture delivers a unified and scalable approach to intelligent automation in \gls{6G} \gls{RAN} environments. The case study demonstrated that this Mamba's \gls{SSM}-driven design significantly improves delay, throughput, energy efficiency, and reliability compared with existing \gls{IDN} management approaches, while reducing computational overhead through efficient sequence modeling. Looking ahead, scaling the proposed architecture to larger deployments or conducting prototype-level validation (e.g., on Open \gls{RAN} compliant testbeds) represents an important direction for future work.
\vspace{-5pt}

\label{s5}

\section*{Acknowledgment}
This work has been supported by MITACS, Ericsson Canada, and NSERC Canada Research Chairs Program.
\vspace{-5pt}

\bibliographystyle{IEEEtran}
\bibliography{Reference.bib}
\end{document}